\documentstyle[psfig]{l-aa}

\def\double {\baselineskip=0.8truecm
             \lineskip=0pt
             \lineskiplimit=0pt}
\def\kms{\,km\,s$^{-1}$}

\def\micron{\,$\mu$m}
 
\def\marc{mag~arcsec$^{-2}$}

\begin{document}

\double
\thesaurus{11.06.2;
	   11.16.1;
	   13.09.1
	   }

\title{1.65~$\rm \bf \mu$m (H-band) surface photometry of galaxies. III: 
observations of 558 galaxies with the TIRGO 1.5m telescope.\thanks{Based on observations taken at TIRGO (Gornergrat, Switzerland). 
TIRGO is operated by CAISMI-CNR, Arcetri, Firenze, Italy.}}


\author{G. Gavazzi\inst{1}
\and P. Franzetti \inst{1}
\and M. Scodeggio \inst{2}
\and A. Boselli\inst{3}
\and D. Pierini\inst{4}  
\and C. Baffa\inst{5} 
\and F. Lisi\inst{5} 
\and L.K. Hunt\inst{6} }

\offprints{G. Gavazzi}

\institute{Universit\`a degli Studi di Milano - Bicocca, P.zza dell'Ateneo Nuovo 1, 20126 Milano, Italy
\and 
European Southern Observatory, Karl-Schwarzschild-Str. 2, D-85748 Garching bei M\"unchen, Germany
\and
Laboratoire d'Astronomie Spatiale, Traverse du Siphon,F-13376 Marseille Cedex 12, France
\and
MPI f\"ur Kernphysik, postfach 103980, D-69117 Heidelberg, Germany
\and
Osservatorio Astrofisico di Arcetri, L.go E.Fermi 5, 50125
Firenze, Italy
\and
C.A.I.S.M.I., L.go E.Fermi 5, 50125 Firenze, Italy }

\date{Received..........; accepted..........}

\maketitle

\markboth{G. Gavazzi et al.: NIR surface photometry of 558 galaxies}{}

\begin{abstract}

We present near-infrared H-band (1.65\micron ) surface photometry of 558
galaxies in the Coma Supercluster and in the Virgo cluster. 
This data set, obtained with the Arcetri NICMOS3 camera ARNICA
mounted on the Gornergrat Infrared Telescope, 
is aimed at
complementing, with observations of mostly early-type objects, our NIR survey
of spiral galaxies in these regions, presented in previous papers of this series.
Magnitudes at the optical radius, total magnitudes, isophotal radii and light
concentration indices are derived. 
We confirm the existence of a positive correlation between the near-infrared 
concentration index and the galaxy H-band luminosity.
\footnote{Tables 1 and 2 are only available in electronic form at
the CDS via anonymous ftp to cdsarc.u-strasbg.fr (130.79.128.5)
or via http://cdsweb.u-strasbg.fr/Abstract.html}

\end{abstract}

\section{Introduction}

Over the last ten years or so, the advent of large format near-infrared (NIR) detectors 
has made images at wavelengths longer than 1\micron\ relatively easy to obtain. 
New infrared cameras based on these detectors make it straightforward to study statistically 
significant samples of galaxies in the NIR (see e.g. de Jong \& van der Kruit 1994). 
The NIR wavelengths constitute the spectral region best adapted to studies of the 
quiescent stellar component of galaxies, since they 
trace mass better than do optical bands, being less contaminated 
by the low M/L products of recent episodes of star formation. 
The NIR spectral region is also relatively free of effects of dust, as
extinction at 1.65 \micron\
is more than seven times lower than in the B band 
(e.g., Landini et al. 1984). 
On the negative side, 
the sky brightness can be as much as ten magnitudes brighter than in the visible; 
thus more sophisticated and time-consuming observing and data reduction 
techniques are required. 

Since 1993 we have used NIR arrays extensively 
to obtain images in the H bandpass of galaxies.
The observing samples were selected by choosing members of the 
Virgo, Coma and A1367 clusters. 
In addition, we selected a significant 
population of galaxies in the portion of the ``Great Wall'' which lies 
in the bridge between Coma and A1367. 
These latter objects can be considered isolated and will be used 
as a control sample for environmental studies. 

Our previous papers concentrated on spiral galaxies. 
About 160 such objects observed with the Calar
Alto 2.2 m telescope were reported in Gavazzi et al. 
(1996c; Paper I); 
another 300, observed similarly with the TIRGO
1.5 m telescope, are given in Gavazzi et al. (1996b; Paper II);
and another
102 Virgo galaxies are reported in Boselli et al. (1997; B97).  
More Calar Alto observations of 170 galaxies in the Virgo cluster
are given in Boselli et al. (1999, Paper IV).
In this paper, we present 1.65\micron\ (H-band) surface brightness 
measurements, obtained in 1997 at TIRGO equipped with the 
Arcetri NIR camera ARNICA, 
of 558 galaxies which are primarily of early-type (E-S0-S0a).
We also provide several repeated measurements
of late-type objects with previously unreliable photometry.
Section 2 of the present paper describes the current sample, and  
the observations are outlined in Section 3.
Image analysis strategies are discussed in Section 4. 
Preliminary results are given in Section 5 and summarized in Section 6.
Profile decompositions using
combinations of exponential profiles and de Vaucouleurs laws of the present data and of
those obtained similarly (Paper I, II, IV and B97) 
will be given in a forthcoming paper (Paper V of this series, Gavazzi et al. 1999a).

\section {Sample selection}

We report on the NIR H band observations of 558 optically-selected 
galaxies, for the most part (457) of early-type morphology, found 
in the regions of the Coma Supercluster and of the Virgo cluster. 
There are
383 early-type galaxies in the CGCG catalogue (m$\rm _p \le$~15.7)
(Zwicky et al. 1961-68) that are members of the Coma supercluster
($\rm 18^o \le \delta \le 32^o$; $\rm 11.5^h \le \alpha \le 13.5^h$),
according to Gavazzi et al. (1999), i.e. with 5000 $<$ V $<$ 8000 \kms.
Of these, we have observed 372 objects, which  
constitute 97\% of the designated sample.   

We also observed 81 of the 98 early-type galaxies brighter than $m_p$=14.0
listed in the VCC catalogue (Binggeli et al. 1985) which are bona-fide members
of the Virgo cluster. 
Thus we have observed 83\% of the giant early-type galaxies in Virgo. 

The remaining 101 observations of spiral galaxies given here
do not form a complete sample in any sense.
However, they are necessary to define a complete data set.
When combined with data published in Papers I and II 
(both of which were devoted to observations of spiral galaxies), 
IV of this series
and B97 (which contains mainly measurements of spiral galaxies in Virgo), 
the NIR survey presented in this series of papers represents a complete, 
optically selected sample of galaxies, which will be analyzed in future papers.

The first 9 columns of Table 1 give the optical parameters of the 558 galaxies studied
in the present work as follows: \newline
Column 1: CGCG (Zwicky et al. 1961-68) or VCC (Binggeli et al. 1985) denomination.  \newline
Column 2: NGC/IC names. \newline
Column 3, 4: adopted (1950) celestial coordinates, measured by us or taken from NED 
\footnote{NASA-IPAC Extragalactic Databasa (NED) is operated by the Jet Propulsion Laboratory,
California Institute of Technology, under contract with NASA},
with few arcsec uncertainty. \newline
Column 5: ``aggregation'' parameter. This parameter defines the membership to a 
group/cluster/supercluster: CSisol, CSpairs, CSgroups indicate members of the Coma Supercluster 
(5000 $<$ V $<$ 8000 \kms); CSforeg means objects in the foreground of the Coma Supercluster 
(V $<$ 5000 \kms) and CSbackg means objects in the background of the Coma Supercluster 
(V $>$ 8000 \kms). Galaxies in the Virgo region are labelled following the membership criteria
given by Binggeli et al. (1993): VCA, VCB, VCM, VCW, VCW', VCSE, VCmem,
are members to the cluster A or B, to the M, W, W' or South-East clouds or are not better specified
members to the Virgo cluster respectively. NOVCC are galaxies taken from the CGCG in the
outskirts of Virgo, but outside the area covered by the VCC. VCback are galaxies in the background
of the Virgo cluster (V$>$3000 km/sec).\newline
Column 6: photographic magnitude as given in the CGCG or in the VCC. \newline
Column 7,8: for CGCG galaxies 
these are the major and minor optical diameters (a$_{25}$, b$_{25}$) (in arcmin) 
derived at the B band $25^{th}$ \marc, as explained in Gavazzi \& Boselli (1996). 
These diameters are consistent with those given in the RC3. 
For VCC galaxies these are
the diameters measured on the du Pont plates at the 
faintest detectable isophote, as listed in the VCC. \newline
Column 9: morphological type. \newline 

\section{Observations}

The observations reported in this paper were acquired with
the 1.5~m, f/20 TIRGO telescope from March 13 to April 13, 1997, when
32 nights were allocated to the present project.
Only 22/32 nights were useful, 16 of which were entirely or partly photometric.
The seeing ranged from 1.5 to 3.5 arcsec (FWHM) with a mean of 2.4 arcsec, 
as shown in Fig. 1. 
These seeing conditions are mostly due to the large pixels ($\sim$\,1\,arcsec)
of ARNICA at TIRGO, and as such represent a necessary disadvantage, because they
also provide the large field-of-view (4.1$\times$4.1 arcmin$^2$)
fundamental for our observations.

\begin{figure}
\psfig{figure=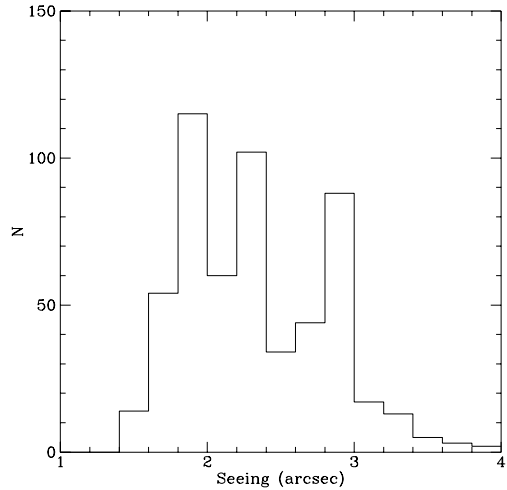,width=10cm,height=10cm}
\caption{The seeing distribution.}
\label{fig.1}
\end{figure}

The folded Cassegrain focus of the telescope is equipped with the
Arcetri NIR camera, ARNICA, which relies on 
a 256$^2$ NICMOS3 array detector (Lisi et al. 1993; 
Lisi et al. 1996; Hunt et al. 1996). 
With a pixel size of 0.97 arcsec, the field-of-view is 4.1$\times$4.1
arcmin$^2$. 
Obtaining a satisfactory background subtraction is the main difficulty
of NIR observations. 
At 1.65\micron, the sky brightness at the Gornergrat is 
$\approx$~13.0\--13.5~\marc, and varies on time scales comparable with the 
typical duration of an observation (e.g., Wainscoat \& Cowie 1992).
To achieve a brightness limit 8~\marc\ fainter than the sky requires an image 
in which the deviations from flatness do not exceed 0.06\%.
Thus, data acquisition techniques must be able to monitor the sky fluctuations 
and data reduction must take these into account. 

To this end,
we used two types of pointing sequences, or ``mosaics'', according to the size of the galaxy. 
Galaxies with an optical diameter $\ge$~1~arcmin were observed
using a sequence in which half of the time is devoted to the target, 
and half to the surrounding sky (hereafter denoted as a type ``A'' 
mosaic)\footnote{Sketches of the "mosaics" used can be found in Fig. 2 of B97.}. 
Typically eight on-source pointings
were alternated with eight on the sky, positioned along a circular 
path around the galaxy and offset by 4~arcmin from the source position. 
The on-source positions were ``dithered'' 
by 10~arcsec in order to facilitate the elimination of bad pixels. 
To save telescope time, small galaxies (with an optical diameter 
$<$~1~arcmin were observed with a pointing sequence consisting 
of nine pointings along a circular 
path and displaced from one another by 1~arcmin 
such that the target galaxy is always in the field 
(hereafter denoted as a type ``B'' mosaic).
 
On-chip integration times were set to 6 seconds to avoid saturation 
but to ensure background-limited performance. 
Total integration times on-source were typically 400~s and ranged from 150 to 950~s. 

Table 1 gives the parameters relevant to the NIR observations as follows:\newline
Column 10: number of frames $N_f$ combined to form the final image (depending on the
adopted mosaic). \newline
Column 11: number of elementary observations (coadds) $N_c$. The total integration time (in seconds) is the product of the number of coadds $N_c$ times the number of combined frames $N_f$ 
times the on-chip integration time $t_{int}$ which was set to 6 sec. \newline
Column 12: seeing (in pixels, with 0.97 arcsec per pixel). \newline
Column 13: adopted zero point (mag / sec).  \newline
Column 14-17: observing dates (day-month-1997); \newline

\begin{figure*}
\psfig{figure=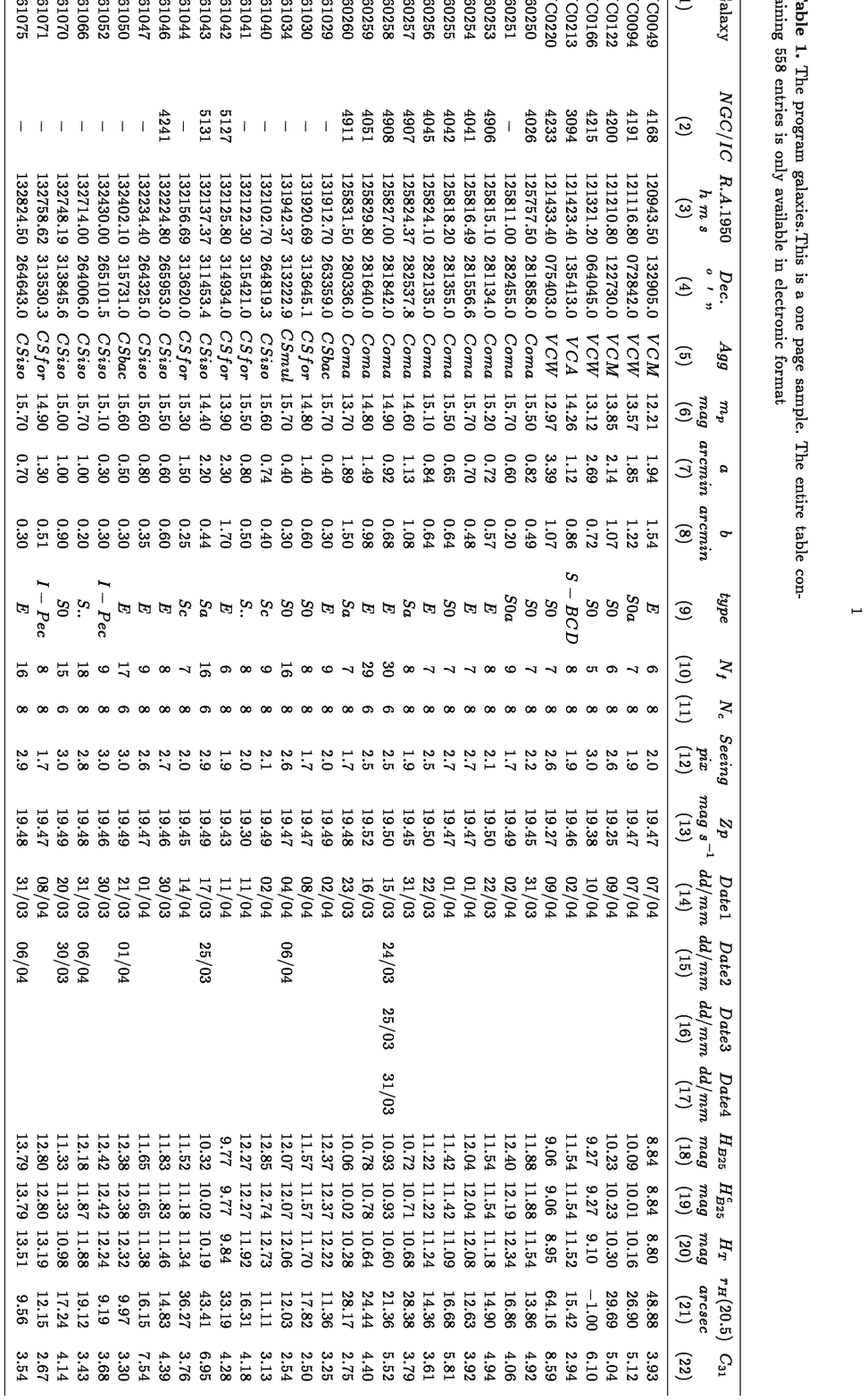,width=18cm,height=23cm}
\end{figure*}

\subsection{Photometric calibration}

Observations of the standard stars HD~84800 (H~=~7.53 mag) and HD~129653
(H~=~6.92 mag) from Elias et al. (1982) were taken hourly throughout the nights for
calibration purposes. 
The calibration stars were observed with a third pointing sequence
(mosaic type ``C'') which consisted of five positions, 
starting with the star near the center of the array, 
followed by positioning the star in each of the four quadrants of the array. 
The telescope was defocussed to avoid saturation. 
During the photometric nights (i.e. those of march 20, 21, 25, 27, 29, 31 and of April
1, 2, 6, 7 and 13) the typical uncertainty in the nightly calibration is 0.05~mag.
Those nights were mostly used to observe galaxies with unknown H band photometry. 

In the remaining non-photometric or marginally photometric nights 
we did not rely on the photometric calibration; 
instead we conservatively observed only those
galaxies with aperture photometry available from the literature.  
For these galaxies the calibration was derived from the 
"virtual aperture photometry" (see Section 4.1 below).

Several (108) galaxies were observed in more than one night 
(see Columns 14-17 in Table 1).
This was done either to check the photometric consistency, 
because some observations were taken in non-photometric conditions, 
or, if the objects were fainter than average, to obtain a longer integration. 
For these objects we obtained an average 
frame by combining the various sets of observations, and using the zero point 
from the frame(s) taken under photometric conditions. 

\subsection{Data reduction procedures}

Since dome exposures cannot be obtained at TIRGO due to the vicinity of 
the telescope secondary ring to the dome, 
the multiplicative correction for the system response, or flat-field 
(FF), was derived daily from the observations.
Many tens (typically 30 to 40) sky frames taken in similar conditions 
throughout the night were combined with a median filter. 
The combined frame, normalized to its median counts 
was used as the flat-field frame. 

The reduction procedure varied according to the type of mosaic. 
For type ``A'' mosaics, the (usually eight) sky exposures were 
combined with a median algorithm to form a median sky. 
For type ``B'' and ``C'' mosaics, the median sky was determined by 
combining all the frames in the pointing sequence.
The median algorithm is necessary to remove unwanted star and 
galaxy images in the median sky frames. 
The median sky was first normalized to its median, 
then multiplied by the  median counts of the individual target frames.
Finally this rescaled frame was subtracted from each of the target observations.
Such a procedure accounts for temporal variations in the sky level 
which are on the order of 5\% during a pointing sequence, 
but introduces an additive offset which is subsequently removed (see below). 
The sky-subtracted target frames are then divided by the FF frame. 
Each of these corrected frames was then analyzed for 
low-spatial-frequency gradients, and if necessary, 
fitted with a two-dimensional 3 degree polynomial 
which was then subtracted.
(We carefully checked that this procedure did not produce artificial 
features which could disturb the photometry of the target objects). 
If this process was not effective in removing the spatial gradients, 
the corresponding frames were rejected from further analysis. 
Finally, the corrected frames were registered using field 
stars and combined with a median filter which allows bad pixel removal. 
Foreground stars were eliminated by manual ``editing'' of the target frames. 
All image reduction and analysis was performed in the IRAF environment 
and relied on the STSDAS package.\footnote{IRAF is the 
Image Analysis and Reduction Facility made available to the astronomical community by the 
National Optical Astronomy Observatories, which are operated by AURA, Inc., under contract 
with the U.S. National Science Foundation. STSDAS is distributed by the Space Telescope Science 
Institute, which is operated by the Association of Universities for Research in Astronomy (AURA), 
Inc., under NASA contract NAS 5--26555.} 

We have assessed the quality of the final images both on small spatial scales, 
and over the entire array. 
All of the images are truly background limited, 
as the noise we measure is the same as that which we would theoretically 
expect from the statistical fluctuations in the sky background, according to
Hunt \& Mannucci (1998) (see Fig. 2). 
The typical pixel to pixel fluctuations are $\sim$~21\marc~, 
i.e. 0.05-0.06\% of the sky. 

\begin{figure}
\psfig{figure=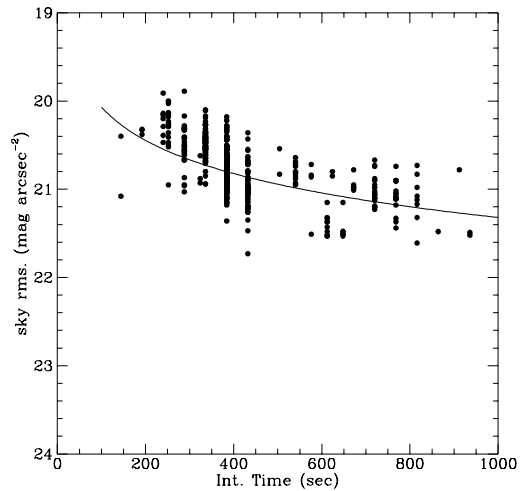,width=10cm,height=10cm}
\caption{The distribution of the sky rms as a function of integration time.
The solid line is the prediction by Hunt \& Mannucci (1998).}
\label{fig.2}
\end{figure}

\section{Image analysis}

\subsection{H-band magnitudes from virtual aperture photometry}

At the position of the galaxies' centers (determined by fitting a two-dimensional
gaussian to the galaxies)
a growth curve was derived for each 
object by integrating the counts in concentric circular rings of increasing radii. 
The background was determined in a concentric object--free corona (where the contribution 
from contaminating stars is rejected by average sigma--clipping).
The measurements taken through the individual "virtual circular apertures" 
are useful to compare with existing similarly taken measurements 
(e.g. Gezari et al. 1993) and for calibration
purposes of future observations. 
They are given in Table 2
(available only in digital format) as follows: \newline
Column 1: Galaxy denomination in the CGCG (Z) or VCC catalogues. \newline
Column 2: aperture diameter in arcsec.\newline
Column 3: logarithmic ratio of the adopted aperture diameter to the optical (a$_{25}$) diameter.\newline
Column 4: integrated H magnitude within the aperture. \newline

\begin{table}
\caption{The "virtual aperture photometry". This is a one page sample. The entire table
containing 6293 entries is only available in electronic format.}
\label{Tab2}
\[
\begin{array}{p{0.15\linewidth}cccc}
\hline
\noalign{\smallskip}

Galaxy & Ap. & log Ap/a_{25} & H  \\ 
      & arcsec &  & mag \\
 (1)  & (2) & (3) & (4) \\
\noalign{\smallskip}
\hline
\noalign{\smallskip}

Z 97005 &   3.80 & -1.15 & 15.31 \\
Z 97005 &   5.80 & -.97 & 14.58 \\
Z 97005 &   7.70 & -.85 & 14.09 \\
Z 97005 &   9.60 & -.75 & 13.78 \\
Z 97005 &  11.50 & -.67 & 13.52 \\
Z 97005 &  15.40 & -.54 & 13.15 \\
Z 97005 &  16.70 & -.51 & 13.05 \\
Z 97005 &  19.20 & -.45 & 12.92 \\
Z 97005 &  22.50 & -.38 & 12.80 \\
Z 97005 &  26.90 & -.30 & 12.67 \\
Z 97005 &  30.70 & -.25 & 12.61 \\
Z 97005 &  34.60 & -.19 & 12.56 \\
Z 97005 &  38.40 & -.15 & 12.53 \\
Z 97011 &   3.80 & -.80 & 14.67 \\
Z 97011 &   5.80 & -.62 & 14.04 \\
Z 97011 &   7.70 & -.49 & 13.67 \\
Z 97011 &   9.60 & -.40 & 13.43 \\
Z 97011 &  11.50 & -.32 & 13.26 \\
Z 97011 &  15.40 & -.19 & 13.09 \\
Z 97011 &  16.70 & -.16 & 13.05 \\
Z 97011 &  19.20 & -.10 & 13.01 \\
Z 97011 &  22.50 & -.03 & 12.97 \\
Z 97011 &  26.90 &  .05 & 12.97 \\
Z 97013 &   3.80 & -1.10 & 16.29 \\
Z 97013 &   5.80 & -.92 & 15.61 \\
Z 97013 &   7.70 & -.79 & 15.21 \\
Z 97013 &   9.60 & -.70 & 14.96 \\
Z 97013 &  11.50 & -.62 & 14.76 \\
Z 97013 &  15.40 & -.49 & 14.58 \\
Z 97013 &  16.70 & -.46 & 14.53 \\
Z 97013 &  19.20 & -.40 & 14.53 \\
Z 97013 &  22.50 & -.33 & 14.51 \\
Z 97021 &   3.80 & -.98 & 12.59 \\
Z 97021 &   5.80 & -.79 & 12.12 \\
Z 97021 &   7.70 & -.67 & 11.87 \\
Z 97021 &   9.60 & -.57 & 11.71 \\
Z 97021 &  11.50 & -.50 & 11.58 \\
Z 97021 &  15.40 & -.37 & 11.41 \\
Z 97021 &  16.70 & -.33 & 11.36 \\
Z 97021 &  19.20 & -.27 & 11.29 \\
Z 97021 &  22.50 & -.20 & 11.21 \\
Z 97021 &  27.10 & -.12 & 11.13 \\
Z 97021 &  31.50 & -.06 & 11.07 \\
Z 97021 &  34.90 & -.01 & 11.02 \\
Z 97021 &  38.40 &  .03 & 10.98 \\
\noalign{\smallskip}
\hline
\end{array}
\]
\end{table}

The photometry of 159 galaxies observed under photometric conditions
has been checked against 187 published aperture photometry measurements
(see Gezari et al. 1993).
The comparison of our "virtual aperture" measurements with the reference photometry, 
taken through apertures consistent with ours, is given in Fig. 3. On
the average we find:

\noindent
$\rm H_{this~work} - H_{literature}$ = -0.026 $\pm$ 0.095 mag.  
\noindent

The most discrepant measurements are those taken through small apertures (5-15 arcsec),
due to a combination of seeing effects and unaccurate galaxy centering. 
We estimate the overall photometric accuracy of our data, 
including systematic errors in the calibration, to be $\leq$~0.1~mag.

\begin{figure}
\psfig{figure=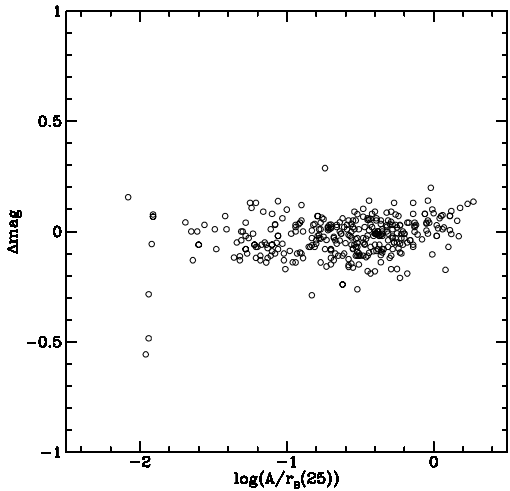,width=10cm,height=10cm}
\caption{The comparison between the present photometric measurements and those available
from the literature as a function of the normalized aperture.}
\label{fig.3}
\end{figure}

\subsection{Isophotal radii, total magnitudes, and concentration indices}

Our NIR observations available so far (including those given in Papers I, II and 
IV and in B97)
were also analyzed using a more sophisticated approach than the one adopted here:
1) the surface brightness profiles were derived from azimuthally averaged elliptical
isophotes with center, ellipticity and position angle taken as free parameters.
2) the surface brightness profiles were decomposed into combinations of exponential
and de Vaucouleurs laws (or double exponential profiles) using a fitting algorithm.
We prefer to postpone the discussion of these profile decompositions to a 
forthcoming paper (Paper V of this series; 
some details of the decomposition procedures can be found in Scodeggio et al. 1998). 
However we wish to anticipate here a few results which depend on the
fitting algorithms only as a tool necessary to extrapolate the measured photometry
to infinity: i.e. to obtain total magnitudes, concentration indexes and also 
to derive isophotal radii at a given limiting surface brightness (20.5 \marc in this case)
when the observations do not reach such a limit. \newline
Table 1 gives the H band measured parameters of the observed galaxies, as follows: \newline
Column 18: $H_{B25}$ magnitude obtained extrapolating the present photometric 
measurements to the optical diameter along circular apertures as in Gavazzi \& Boselli (1996). \newline
Column 19: $H_{B25}^c$ magnitude computed at the optical diameter (see Column 18) corrected
for galactic and internal extinction following Gavazzi \& Boselli (1996). 
The adopted internal 
extinction correction is $\rm \Delta m=-2.5 \, D\, \log(b/a)$ where D=0.17, as determined 
in Boselli \& Gavazzi (1994).\newline
Column 20: $H_T$ total H magnitude extrapolated to infinity along either an
exponential or a de Vaucouleurs $r^{1/4}$ law fitted to the outer parts of the 
observed radial surface brightness profiles (see Paper V for details). \newline
Column 21: galaxy observed major ($r_H(20.5)$) radius (in arcsec) determined 
in the elliptical azimuthally--integrated 
profiles as the radius at which the surface
brightness reaches 20.5 H--\marc. 
Galaxies which require a surface brightness extrapolation larger than
0.5 mag to reach the $\rm 20.5^{th}$ ~magnitude isophote are labelled -1. \newline
Column 22: the model--independent concentration index $\rm C_{31}$,
as defined in de Vaucouleurs (1977), is the ratio between the radii that 
enclose 75\% and 25\% of the total light $H_T$. \newline

\section{Results}

Based on the observations presented in this work, which we reiterate does not comprise 
a complete sample, we derive the following preliminary results.

\subsection{Radii $r_H(20.5)$}

The isophotal radii in this work are derived at the 20.5 \marc~H-band isophote, which
represents a rather bright level, even for H-band measurements. For example in Paper I and II
we were able to measure similar quantities one \marc~fainter, i.e. up to 21.5 \marc. 
This is not due to lower signal-to-noise of the present data compared with the past data,
but rather to the different method adopted for deriving the light profiles. In the previous
papers the radial surface brightness profiles were derived by azimuthally integrating
the counts in concentric elliptical coronae of fixed center and ellipticity up to
indefinite radii. 
Here, instead, the center and ellipticity are kept as free parameters and
the fitting is halted when the mean surface brightness
within a given elliptical corona equals the corresponding rms. fluctuation. 
The drawback is that the new fitting routine halts at higher surface brightnesses than before.  
The lowest surface brightness reached in each image is given in Fig. 4 as a function of the integration time.

\begin{figure}
\psfig{figure=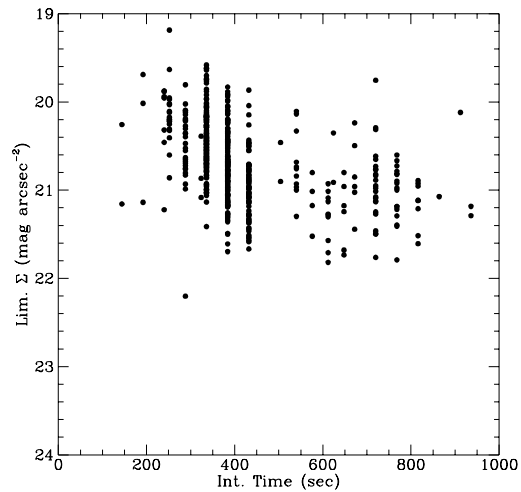,width=10cm,height=10cm}
\caption{The distribution of the lowest surface brightness reached in the outer light profiles,
as a function of the exposure time.}
\label{fig.4}
\end{figure}

The comparison between the isophotal B band $r_B(25.0)=(a/2)_{25}$(B) radii
(Gavazzi \& Boselli 1996) and the infrared $r_H(20.5)$ isophotal radii
determined in this work is shown in Fig. 5. Apart from the 
curvature apparent at small radii, the data are consistent with 
$r_H(20.5)~=~0.7~r_B(25.0)$ 
between these two arbitrary isophotal levels. 

This indicates that at the adopted isophotal level, the H  
observations cover a substantial fraction of the light, and
are not restricted to the bulge.
The relationship between optical and H-band isophotal radii is consistent
with that expected from our limiting H surface brightness of 20.5 \marc\ and
with the B--H color of the outer portion of normal galaxies.

The curvature seen in Fig. 5 is only marginally an artefact of the seeing: 
the effect is reduced slightly by removing the objects observed in the 
worse seeing conditions.
The deviation from linearity is not due to an overestimate of the H band radii,
rather it reflects an underestimate of the B band diameter
of small ($\leq~10$ arcsec) early-type galaxies which are intrinsically red
(see Paper V for a more comprehensive discussion on this issue).

\begin{figure}
\psfig{figure=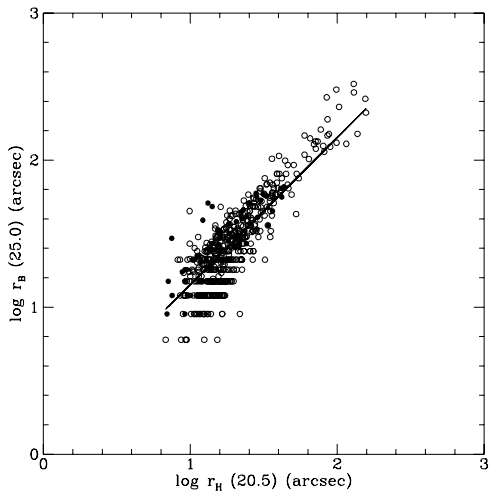,width=10cm,height=10cm}
\caption{The relation between the apparent major isophotal radius $r_H(20.5)$  
as determined in the infrared (this work) and in the optical $r_B(25.0)$.
E+S0 are plotted as open circles, S+Irr as filled circles.
The solid line represents the relation $r_H(20.5)~=~0.7~r_B(25.0)$.}
\label{fig.5}
\end{figure}

\subsection{Magnitudes ($H_T$, $H_{B25}$)}

$H_{B25}$ magnitudes listed in Column 18 of Table 1 are obtained by extrapolating 
the circular aperture measurements to the optical $r_B(25.0)$ radius
(as in Gavazzi \& Boselli 1996). 
$H_T$ mag instead are obtained by extrapolating to infinity the magnitude 
integrated along elliptical isophotes using combinations of exponential and 
de Vaucouleurs laws. 
As expected, $H_T$ are brighter than H$_{B25}$ by $0.10\pm0.2$~mag on average.

\subsection{Concentration index ($C_{31}$)}

The concentration index $C_{31}$ is a 
measure of the shape of light profiles in galaxies, independent of
a bulge--disk decomposition. 
Values larger than $C_{31}\,>\,2.8$ indicate the presence of
substantial bulges.

We confirm the presence in our sample of a general
correlation between $C_{31}$ and the H band ($H_T$ or $H_{B25}$) luminosity
(computed from the redshift distance). 
We find that $C_{31}$ generally increases toward 
higher absolute magnitudes (Fig. 6).
High $C_{31}$ are found only among high luminosity systems, 
but the converse is not true:
there are several high luminosity systems 
(namely late--type giant spirals) with no or little bulge
($C_{31}\sim 3$).

\begin{figure}
\psfig{figure=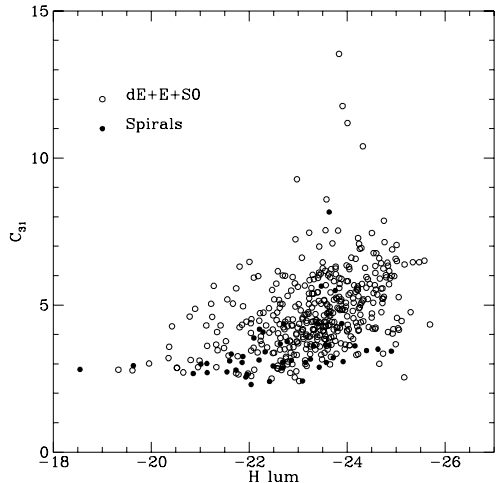,width=10cm,height=10cm}
\caption{The dependence of the near--infrared concentration index $C_{31}$ on H band
luminosity.}
\label{fig.6}
\end{figure}

\section{Summary}

We have obtained images in the near-infrared H bandpass for
an optical-magnitude-selected ($m_p \le 15.7$) sample of 558 nearby ($z<0.02$) galaxies
primarily of early-type, in the regions of Coma and Virgo. 
We derived H magnitudes at the optical radius, total H magnitudes, isophotal
radii at the 20.5 \marc~ isophote, and the light concentration index $C_{31}$.
As mentioned in the Introduction, the galaxies presented in this
paper do not form by themselves a complete sample. 
Therefore we postpone the comprehensive analysis of the NIR properties
of galaxies to forthcoming papers of this series. Paper V will report on
the profile decomposition.

\acknowledgements

We wish to thank the TIRGO T.A.C. for the generous time allocation to this project 
and the TIRGO team for support during the observations.
We thank A. Borriello,  V. Calamai, B. Catinella, I. Randone, 
P. Ranfagni, M. Sozzi, P. Strambio for assistance during the
observations at TIRGO. A special thanks to V. Gavriusev for software
assistance at TIRGO.

\end{document}